\documentclass[%
superscriptaddress,
preprint,
 amsmath,amssymb,
 aps,
]{revtex4-1}

\usepackage{graphicx}
\usepackage{dcolumn}
\usepackage{bm}

\usepackage[colorlinks=true,linkcolor=blue,citecolor=blue]{hyperref}

\begin{document}


\title{Bubble nucleation around heterogeneities in $\phi^4$-field theories}

\author{\textcopyright Juan F. Mar\'in
}
 \email{juan.marin.m@usach.cl}
\affiliation{Departamento de F\'isica, Universidad de Santiago de Chile, Usach, Av. Ecuador 3493, Estaci\'on Central, Santiago, Chile.}%




\date{\today}

\begin{abstract}
Localised heterogeneities have been recently discovered to act as bubble-nucleation sites in nonlinear field theories. Vacuum decay seeded by black holes is one of the most remarkable applications. This article proposes a simple and exactly solvable $\phi^4$  model exhibiting bubble nucleation around localised heterogeneities. Bubbles with a rich dynamical behaviour are observed depending on the topological properties of the heterogeneity. The linear stability analysis of soliton-bubbles predicts the formation of oscillating bubbles and the insertion of new bubbles inside an expanding precursor bubble. Numerical simulations in 2+1 dimensions are in good agreement with theoretical predictions.
\end{abstract}

\maketitle


\section{\label{sec:level1}Introduction}

The role of heterogeneities in scalar field theories is relevant in many physical phenomena, ranging from the stability of the Higgs vacuum to phase transitions in the early Universe \cite{Gregory2014, Burda2015, Burda2015-2, Burda2016}. One of the most intriguing aspects of our current understanding of the Universe is the possibility that the Universe can be trapped in a meta-stable state of the Higgs potential \cite{Turner1982}. From such a meta-stable state --\emph{the false vacuum state}-- a phase transition may occur towards a global minimum--\emph{the true vacuum state}. A transition of the universe towards a nearby global minimum of the Higgs potential-- a phenomenon termed as \emph{vacuum decay} \cite{Weinberg2012},  would lead the Universe towards a state where life as we know it might be impossible \cite{Kibble1980, Turner1982}. The possibility that heterogeneities, such as black holes, may enhance the nucleation of true-vacuum bubbles in the Universe and thus seeding vacuum decay is a fascinating subject that is gaining increasing interest in gravitation and cosmology \cite{Moss1985, Hiscock1987, Berezin1991, Gregory2014, Burda2015, Burda2016, Gonzalez2018, Gregory2019, Cuspinera2019}.

The idea of modelling a black hole as heterogeneity in scalar field theories has been recently proposed by Gregory and co-workers \cite{Gregory2014, Burda2015, Burda2016}. The fundamental mechanism is that black holes emit high energy particles that perturbs any scalar field $\phi$ coupled to the emitted quanta \cite{Cheung2014}. In scalar field theories where $\phi$ resides in a meta-stable state, such perturbation can drive the field $\phi$ over the potential barrier and activate vacuum decay. This scenario corresponds to the nucleation of true-vacuum bubbles around the black hole. For instance, exploding or evaporating black holes would be surrounded by phase-transition bubbles \cite{Moss1985, Nach2019}. This phenomenon is reminiscent to the intuitive picture of the first-order phase transition occurring in boiling water, where bubbles of the new phase are nucleated --often around impurities--, and eventually expands. Other works address the validity of these ideas in the Higgs model, including the effects of gravity \cite{Burda2015}, large extra dimensions \cite{Cuspinera2019}, and deviations from the thin-wall approximation \cite{Burda2016}.

This article introduces an exactly solvable heterogeneous $\phi^4$ field theory exhibiting bubble nucleation around the heterogeneity. It is shown how space-dependent heterogeneities deform the energy landscape of the $\phi^4$ system affecting the stability of the vacua. The topological properties of the heterogeneity are described through a single parameter, and as will be shown, a rich variety of bubble dynamics can be observed according to its value.  Bubbles can be nucleated, oscillate, or even expand. Moreover, it is shown how the interaction of the bubble with the heterogeneity may trigger nonlinear instabilities that may enhance vacuum decay and the insertion of a new bubble inside a precursor bubble.

The outline of the article is as follows. Section \ref{Sec:Bubbles} introduces the nonlinear field theory and the solitonic-model of bubbles. Section \ref{Sec:LinearStability} summarises the results from the linear stability analysis of the soliton-bubbles under the influence of the heterogeneity. The first implication from the linear stability analysis, which is the existence of oscillating bubbles, is discussed in section \ref{Sec:OscillatingStates}. Section \ref{Sec:HeterogeneousVacDec} is devoted to conditions where non-expanding true-vacuum bubbles can be sustained by the heterogeneity. For some combinations of parameters,  vacuum decay can occur. The nucleation of bubbles inside a precursor expanding-bubble is demonstrated in Section \ref{Sec:Precursor}. The robustness of the reported phenomena under thermal fluctuations is briefly discussed in Section \ref{Sec:Noise}. Finally, concluding remarks are given in Section \ref{Sec:Conclusions}.

\section{Bubbles and topological defects in $\phi^4$ systems \label{Sec:Bubbles}}

\begin{figure}
    \centering
    \scalebox{0.7}{\includegraphics{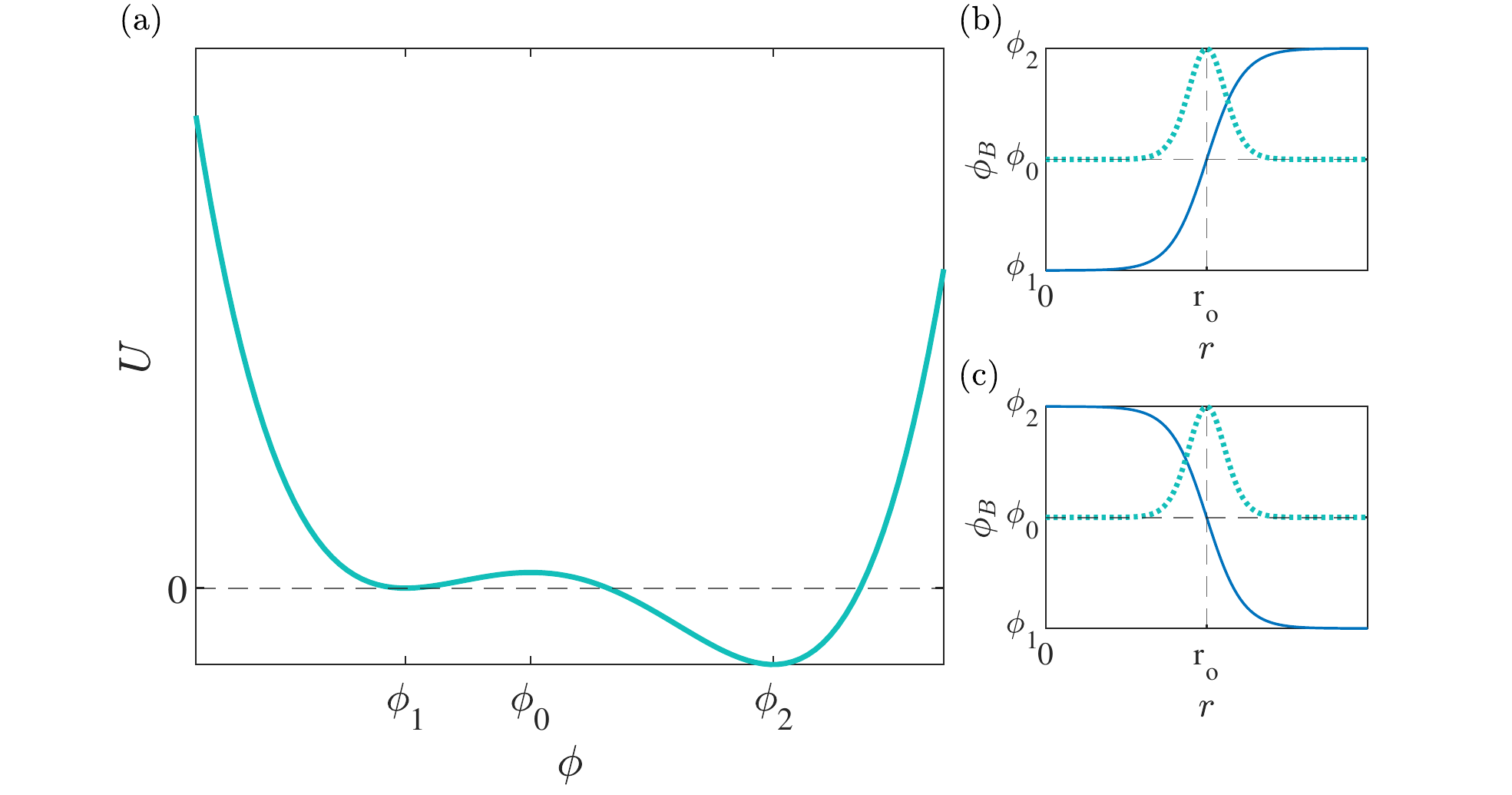}}
    \caption{\textbf{(a)} Effective potential of Eq.~\eqref{Eq:EffectivePotential} for $\mu=1/2$, $\eta=1$ and $f=0.1$ with a barrier at $\phi=\phi_0$ separating a false-vacuum ($\phi=\phi_1$) from a true-vacuum ($\phi=\phi_2$). \textbf{(b)} A soliton bubble of radius $r_o$ representing a bubble of phase $\phi_1$ (a \textit{$\phi_1$-bubble}) in a space filled with phase $\phi_2$. \textbf{(c)} A soliton bubble of radius $r_o$ representing a bubble of phase $\phi_2$ (a \textit{$\phi_2$-bubble}) in a space filled with phase $\phi_1$. The normalised energy density of each bubble is shown in dashed lines.}
    \label{fig:01}
\end{figure}

Bubbles in scalar field theories are configurations where the field $\phi$ is in one phase in a space filled with another phase. In the context of the qualitative theory of nonlinear dynamical systems, bubbles are coexistent heteroclinic trajectories -- or kinks, also known as \textit{topological defects} --joining fixed points of the underlying Higgs potential \cite{Guckenheimer1986}. Ring solitons \cite{Christiansen1981}, kink-antikink pairs \cite{Gonzalez2006}, and other topologically equivalent solutions \cite{Castro-Montes2020} are physically relevant models of the so-called \textit{soliton-bubbles} of the nonlinear field theory. Consider a field theory in $D+1$ dimensions with a single real scalar field $\phi$, whose action is
\begin{equation}
    \label{Eq:01}
    S=\int dt\,d^{D}r\left[\frac{1}{2}\left(\partial_{\alpha}\phi\right)^2-\frac{\mu}{4}(\phi^2-\eta)^2+f\phi\right],
\end{equation}
where $\alpha=0,1,\ldots,D$, $f=f(\mathbf{r})$ is a heterogeneous perturbation, $\mathbf{r}\in\mathbb{R}^D$ and $\mu$ and $\eta$ are parameters of the model. The theory turns into the celebrated $\phi^4$ model for $f=0$, with two degenerate vacua at $\phi=\pm\eta$ separated by a barrier at $\phi=0$ \cite{Vachaspati2006}. If $|f|<1/2\,\forall\mathbf{r}$, the vacua of the effective potential
\begin{equation}
 \label{Eq:EffectivePotential}
 U_{\text{eff}}(\phi)=\frac{\mu}{4}\left(\phi^2-\eta^2\right)^2-f\phi,
\end{equation}
becomes non-degenerate with a true- and a false vacuum state, as depicted in Fig.~\ref{fig:01}(a) for a fixed $\mathbf{r}$. The effect of damping can be taken into account using the phenomenological Rayleigh dissipation function $\mathcal{R}=(\gamma/2)(\partial_t\phi)^2$, where $\gamma$ is the damping constant. The resulting equation of motion is the driven and damped $\phi^4$ equation
\begin{equation}
    \label{Eq:02}
 \partial_{tt}\phi-\nabla^2\phi+\gamma\partial_t\phi-\mu\left(\eta^2-\phi^2\right)\phi=f(\mathbf{r}).
\end{equation}

Topological defects are heteroclinic solutions interpolating the stable phases of the underlying potential \cite{Guckenheimer1986}, namely $\phi_1$ and $\phi_2$. For $D=1$ they are known as kinks \cite{Peyrard2004, Manton2004}, whereas for $D>1$ they are known as \emph{line solitons} or \emph{domain walls} \cite{Christiansen1981, Vachaspati2006}. Let $\phi_1<\phi_2$ hereon. Soliton bubble solutions of Eq.~\eqref{Eq:02} are real and rotationally symmetric solutions, i.e.  $\phi_B(\mathbf{r})=\phi_B(r)$ where $r=(x_1^2+\ldots+x_D^2)^{1/2}$, with the following properties \cite{Barashenkov1989, Barashenkov1993}. A $\phi_1$\emph{-bubble} is a bubble of phase $\phi_1$ in a space filled with phase $\phi_2$ [see Fig.~\ref{fig:01}(b)] in the form of a rotationally symmetric solution $\phi_B(r)$ such that $\lim_{r\to\infty}\phi_B(r)=\phi_2$ and
\begin{subequations}
    \begin{equation}
    \label{Eq:03a}
    0\leq\partial_r\phi_B(0)<\partial_r\phi_B(r)<\infty,
    \end{equation}
    \begin{equation}
    \label{Eq:03b}
        \phi_1 <\phi_B(0)<\phi_B(r)<\phi_2,
    \end{equation}
\end{subequations}
for $r\in(0,\,\infty)$. Similarly, a $\phi_2$\emph{-bubble} is a bubble of phase $\phi_2$ in a space filled with phase $\phi_1$ [see Fig.~\ref{fig:01}(c)] such that $\lim_{r\to\infty}\phi_B(r)=\phi_1$ and 
\begin{subequations}
    \begin{equation}
    \label{Eq:04a}
    0\leq\partial_r\phi_B(0)>\partial_r\phi_B(r)>-\infty,
    \end{equation}
    \begin{equation}
    \label{Eq:04b}
        \phi_2 >\phi_B(0)>\phi_B(r)>\phi_1,
    \end{equation}
\end{subequations}
for $r\in(0,\,\infty)$. 

Phenomenological models with exact or approximate analytical solutions are useful to understand physical phenomena in real systems. Some examples are the SSH theory of nonlinear excitations in polymer chains \cite{Heeger1988} and the Peyrard-Bishop model of denaturation of the DNA \cite{Peyrard1989}, both widely used to answer important questions of chemical and biological interest \cite{Peyrard2004}. In our system, solving an inverse problem, it is possible to give systems of type \eqref{Eq:02} with exact solutions.  Later, in the sense of the qualitative theory of nonlinear dynamical systems \cite{Guckenheimer1986}, it is possible to generalise analytical results obtained for systems of type (\ref{Eq:02}) to other solutions that are topologically equivalent \cite{Gonzalez1996, Gonzalez1999, Gonzalez2003, GarciaNustes2017, Marin2018, Castro-Montes2020}. For instance, the following soliton-bubble
\begin{equation}
\label{Eq:05}
\phi_B(r)=\tanh[B(r-r_o)],
\end{equation}
is an exact solution to Eq.~\eqref{Eq:02} for $\mu=1/2$, $\eta=1$, and
\begin{equation}
\label{Eq:06}
f(\mathbf{r})=\left[\frac{1}{2}(4B^2-1)\tanh[B(r-r_o)]-\frac{(D-1)B}{r}\right]\mbox{sech}^2[B(r-r_o)],
\end{equation}
where $r_o$ is the radius of the bubble and $B$ is a parameter that describes the topological properties of both the bubble and the heterogeneity. Parameter $B$ determines the extreme values of $f(\mathbf{r})$, its decay length, and the value of its derivative at its first-order zero. The density energy of the $\phi_1$-bubble of Eq.~\eqref{Eq:05} is localised at the wall of the bubble [see figures \ref{fig:01}(b) and \ref{fig:01}(c)], forming a disk of radius $r_o$ centred at the origin with decay length $\ell:=1/B$. For small (large) values of parameter $B$, the energy becomes localised in a disk with large (small) width.

\begin{figure}
    \centering
    \scalebox{0.55}{\includegraphics{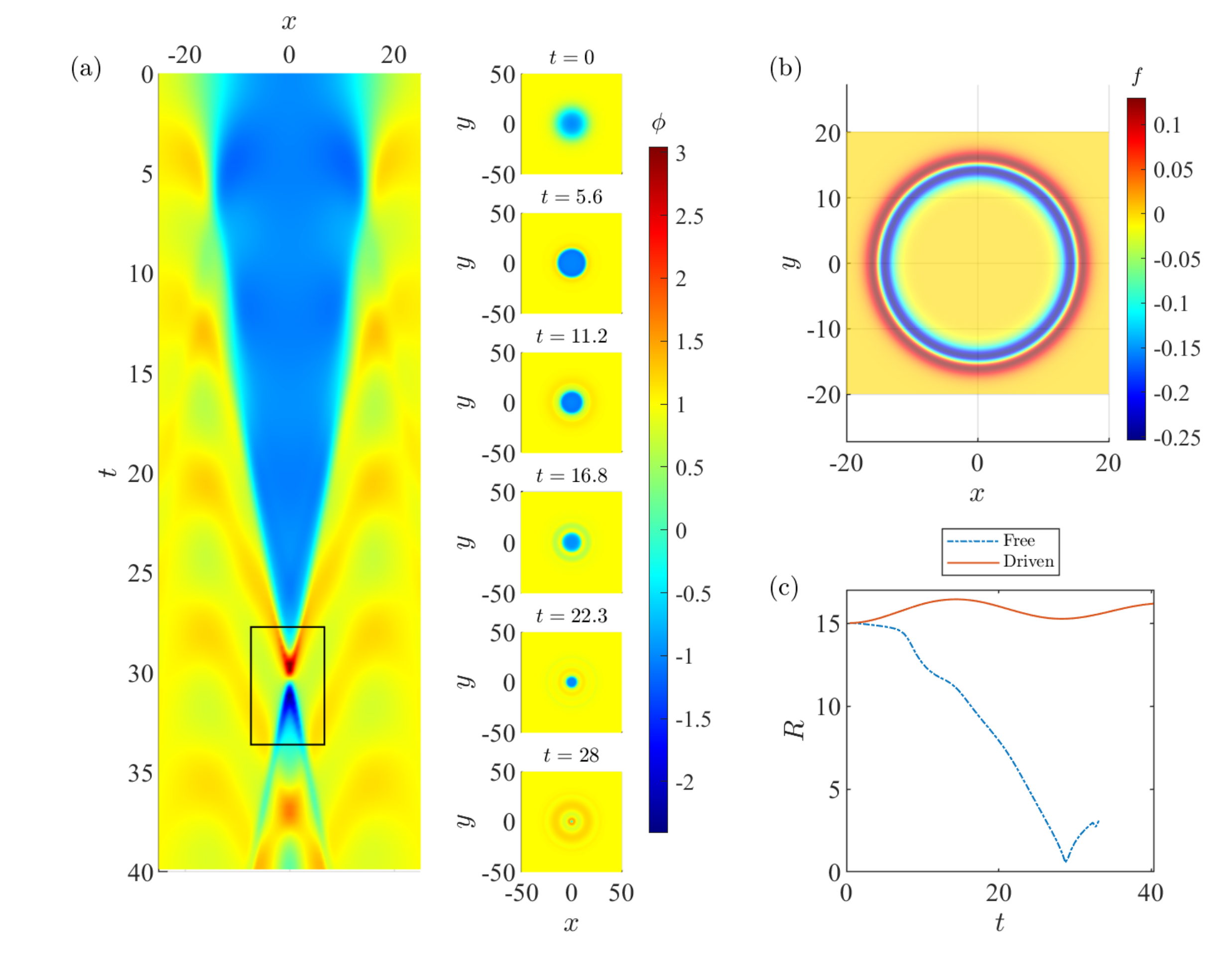}}
    \caption{\textbf{(a)} Collapse of a free bubble of radius $r_o=15$ with $\mu=1/2$, $\eta=1$, $\gamma=0.01$, $B=1$ and $f=0$. The insets show the snapshots of the bubble as a function of space for the indicated values of the time. The localised burst appearing after the collapse is highlighted with a box. \textbf{(b)} The heterogeneous force of Eq.~\eqref{Eq:06} as a function of space for $B=0.6$, exhibiting the shape of a double ring. \textbf{(c)} The radius $R(t)$ for a free bubble (blue dashed-dotted line) and a driven bubble (red solid line) as a function of time, for $r_o=15$, $\mu=1/2$, $\eta=1$, and $\gamma=0.01$. The results for the driven and free bubbles corresponds to $B=0.6$ and $B=1$, respectively. }
    \label{fig:02}
\end{figure}

Numerical simulations of Eq.~\eqref{Eq:02} with the soliton bubble of Eq.~\eqref{Eq:05} as initial condition demonstrates that bubbles always collapse towards its centre if $f=0$. Indeed, the curvature of the bubble wall is proportional to the surface tension of the bubble, which under no other restrictions enhances its collapse. Figure~\ref{fig:02} shows the results from numerical simulation for $D=2$, using homogeneous Neuman boundary conditions for the given values of the parameters. Let $x_1:=x$ and $x_2:=y$ be the space-coordinates in $\mathbb{R}^2$. For all the numerical simulations shown in this article, the time integration was performed using a fourth-order Runge-Kutta scheme with $dt=0.001$ and finite differences of second-order accuracy with $dx=dy=0.25$ for the Laplace operator.

Figure~\ref{fig:02}(a) shows the time evolution of the collapsing bubble profile at $y=0$. The bubble area decreases in time and the collapse is completed for $t=28.73$. Notice that the area of the bubble describes an oscillating behaviour during its collapse, introducing radiative waves that propagate towards the boundaries. These oscillations can be understood from the theory of linear response, studying the excitation spectra of the soliton bubble \eqref{Eq:05}. Similarly to one-dimensional $\phi^4$ kinks, the bubble spectrum of the unperturbed $\phi^4$ equation is composed of a continuous spectrum and a discrete spectrum \cite{Peyrard2004}. The discrete spectrum features a localised translational mode --an extension of the notion of the \textit{Goldstone mode} in translationally-invariant systems. In the gap between the origin of the spectral plane and the continuous spectrum, there is also an internal (localised) shape mode \cite{Peyrard1989, Chirilus-Bruckner2014, Kevrekidis2019}. Such internal mode is responsible for the oscillations in the bubble width, which are nonlinearly coupled to the scattering (phonon) modes in the continuous spectrum. The resulting radiative waves are observed as small-amplitude rings with an increasing radius in time, as observed in the insets of Fig.~\ref{fig:02}(a). These waves eventually decay in the presence of linear damping. The simulation also reveals a space-time localised burst just after the high-energetic collapse of the bubble, which is indicated with a box in Fig.~\ref{fig:02}(a). An emergent small bubble appears after the burst. However, it is short-lived and eventually collapses.

The heterogeneous force of Eq.~\eqref{Eq:06} is depicted in Fig.~\eqref{fig:02}(b), and has the shape of a double ring: a negative ring (or trench) with mean radius $r_*-\ell$ and a positive ring (or hill)  with mean radius $r_*+\ell$, where $r_*$ is the zero of the force. Between both rings, there is a circumference of radius $r_*$ where the force vanishes. Similarly to one-dimensional $\phi^4$ kinks \cite{Gonzalez1992}, the first-order zeros of $f(r)$ are equilibrium positions where the bubble wall can be trapped by the heterogeneity. For a $\phi_1$-bubble ($\phi_2$-bubble) there is a simple condition of the equilibrium position of its wall at $r=r_*$, namely,
\begin{equation}
    \label{Eq:07}
    \left.\frac{df(r)}{dr}\right|_{r=r_*}\underset{(<)}{\displaystyle >}\,0,
\end{equation}
provided that there are no more first-order zeros in the region $[r_*-\ell,r_*+\ell]$. The sign of the derivative in Eq.~\eqref{Eq:07} depends on the value of parameter $B$. In Fig.~\ref{fig:02}(b), the region $r=r_*$ is stable for the wall of a bubble of phase $\phi=-1$ inside a space filled with $\phi=1$. Thus, such bubbles may be sustained by the heterogeneity preventing its collapse. This observation is confirmed in Fig.~\ref{fig:02}(c), where the radius of the bubble, $R(t)$, is depicted as a function of time for both cases, in the presence and the absence of the heterogeneity. The evolution of the radius of the bubble of Eq.~\eqref{Eq:05} under the heterogeneous force of Eq.~\eqref{Eq:06} with $B=0.6$ is shown in solid (red) line in Fig.~\ref{fig:02}(c). Notice that the radius of the bubble performs damped oscillations around the stable equilibrium radius $r_*>r_0$.

\begin{figure}
    \centering
    \scalebox{0.4}{\includegraphics{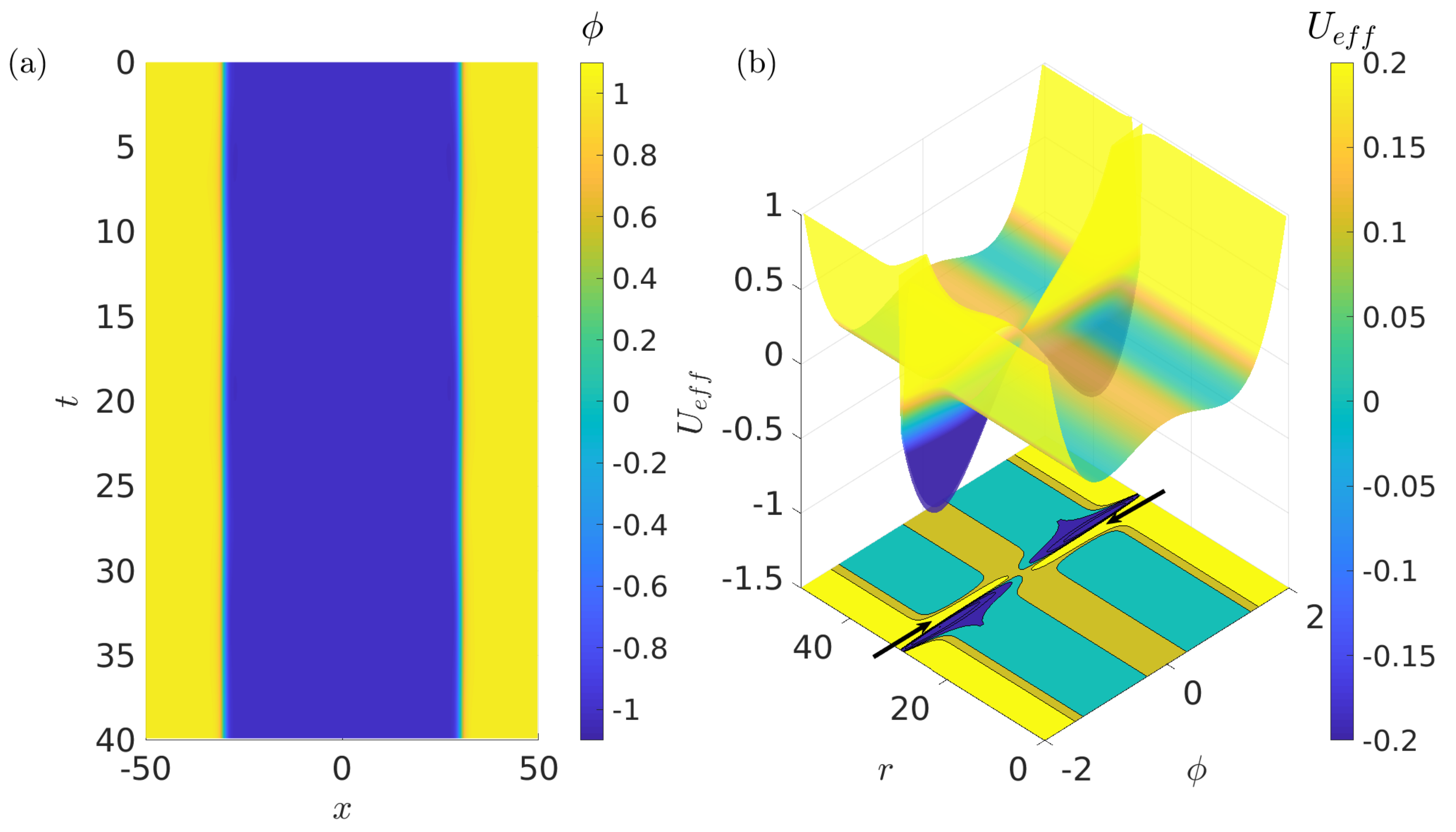}}
    \caption{A stationary bubble for $B=1$, $r_o=30$, $\mu=1/2$, $\eta=1$ and $\gamma=0.01$. \textbf{(a)} Spatiotemporal diagram of the profile of $\phi(x,y,t)$ for $y=0$. Any other profile will exhibit a similar behaviour due to the symmetry of the system. \textbf{(b)} Energy landscape of the system with two slightly non-degenerated minima at $(\phi_1, r_o)$ and $(\phi_2, r_o)$.}
    \label{fig:03}
\end{figure}

If the initial radius of the bubble coincides with the first-order zero of the heterogeneity, i.e. $r_o=r_*$, there will be no damped oscillations and the bubble will remain stationary. This is the case in the numerical simulations shown in Fig.~\ref{fig:03}(a), where a stable and stationary bubble is observed for the given combination of parameters. The effective potential of the system can be written as
$U_{\text{eff}}(\phi)=U(\phi)+\phi f(r)$, where $U(\phi):=(\phi^2-1)^2/8$ is the $\phi^4$ potential in its standard form. Figure \ref{fig:03}(b) shows the effective potential for the same combination of parameters as Fig.~\ref{fig:03}(a). The energy landscape has two slightly non-degenerated minima at the phases $\phi_1<0$ and $\phi_2>0$, both with a radius $r=r_o=30.0$. Notice that both potential wells are relatively narrow and symmetrically placed under reflections with respect to the lines $r=30$ and $\phi=0$. Small perturbations of this bubble configuration will produce damped oscillations inside each potential well. However, the extended bubble-structure will be stable. Notice that there is a barrier --indicated with arrows in the contour diagram at the bottom of Fig.~\ref{fig:03}(b)-- that prevents the phase $\phi_1$ to invade the phase $\phi_2$ throughout all space.

In summary, although the soliton bubble of Eq.~\eqref{Eq:05} collapses when left unperturbed, the presence of the heterogeneity may sustain the bubble and prevent its collapse, enlarging its lifetime. The heterogeneity of Eq.~\eqref{Eq:06} has a singularity at the origin and is able to seed phase transitions, as will be shown later in this article.

\section{Linear stability analysis \label{Sec:LinearStability}}

The effect of heterogeneities and external perturbations in the dynamics of the bubble soliton of Eq.~\eqref{Eq:05} can be investigated performing the linear stability analysis \cite{Peyrard2004}. With this purpose, the driven-damped $\phi^4$ equation (\ref{Eq:02}) can be written as
\begin{equation}
\label{Eq:09}
 \partial_{tt}\phi-\nabla^2\phi+\gamma\partial_t\phi-G(\phi)=f(\mathbf{r}),
\end{equation}
where $G(\phi):=-dU/d\phi$. Consider a small-amplitude perturbation $\chi$ around the soliton-bubble solution $\phi_B$, i.e.
\begin{subequations}
\label{Eq:10}
\begin{equation}
 \label{Eq:10a}
 \phi(\mathbf{r},t)= \phi_B(r)+\chi(r,t),
 \end{equation}
 \begin{equation}
 \label{Eq:10b}
 \chi(r,t):=\psi(r)e^{\lambda t}\quad,\quad |\chi|\ll|\phi_B|\forall(\mathbf{r},t),
\end{equation}
\end{subequations}
with $\lambda\in\mathbb{C}$. Expanding $G(\phi)$ with $\phi\sim\phi_B$, one obtains
\begin{equation}
\label{Eq:11}
    G=\frac{1}{2}\left[\phi_B(1-\phi_B^2)+\chi(1-3\phi_B^2)\right]+\mathcal{O}(\chi^2)\,,\quad\phi\sim\phi_B.
\end{equation}

After substitution of Eqs.~(\ref{Eq:10a}) and~(\ref{Eq:11}) in the $\phi^4$ equation~(\ref{Eq:09}), one obtains for $\psi(\mathbf{r})$ the following  spectral problem
\begin{subequations}
\label{Eq:12}
 \begin{equation}
  \label{Eq:12a}
  \hat{L}\psi=\Gamma \psi,
  \end{equation}
 \begin{equation}
   \label{Eq:12b}
   \Gamma:=-\lambda(\lambda+\gamma),
\end{equation}
\end{subequations}
where the linear operator $\hat{L}$ is
\begin{subequations}
\label{Eq:13}
 \begin{equation}
 \label{Eq:13a}
    \hat{L}:=-\nabla^2+V(r,\phi=\phi_B),
\end{equation}
\begin{equation}
 \label{Eq:13b}
    V(r,\phi=\phi_B)=\left.\frac{d^2U}{d\phi^2}\right|_{\phi=\phi_B}.
\end{equation}
\end{subequations}

Consider the dynamics of bubbles whose shape are far from collapse. Such bubbles can be stable and preserve their shape in time, or rather become unstable due to some nonlinear instability. For almost-collapsed bubbles near its centre, the condition $Br_o\ll 1$ is fulfilled, whereas bubbles far from collapse correspond to the condition $Br_o\gg1$. In the later limit, $r_*\to r_0$, $\nabla^2\simeq d^2/dr^2$ and Eq.~\eqref{Eq:12a} is analogous to the one-dimensional Schr\"odinger equation for a particle in the P\"oschl-Teller potential,
\begin{equation}
    \label{Eq:14}
    V(r)=-\frac{1}{2}\left(1-3\tanh^2[B(r-r_o)]\right).
\end{equation}

Thus, the soliton bubble behaves like a potential well for linear waves. The eigenvalues and eigenfunctions of the spectral problem~\eqref{Eq:12}--\eqref{Eq:14} can be obtained exactly \cite{Flugge2012, Holyst1991, Gonzalez1992, Gonzlez2008}. The discrete spectrum corresponds to soliton-phonon bound states, whose eigenvalues are given by
\begin{subequations}
 \begin{equation}
 \label{Eq:15a}
 \Gamma_n=B^2(\Lambda+2\Lambda n-n^2)-\frac{1}{2},\quad (n=0,\,1,\ldots,[\Lambda]-1),
 \end{equation}
 \begin{equation}
 \label{Eq:15b}
 \text{with }\Lambda(\Lambda+1)=\frac{3}{2B^2},
\end{equation}
\end{subequations}
where $[\Lambda]$ is the integer part of $\Lambda$. Thus, for a fixed value of $B$, the $n$-th bound state exists if
\begin{equation}
\label{Eq:16}
    n<\Lambda.
\end{equation}
This includes the translational mode $\Gamma_o$ and the internal shape modes $\Gamma_n$ with $n>0$. 

To determine the stability condition for a generic $n$-th mode, one has to find conditions for which the real part of $\lambda$ crosses the imaginary axis. From Eq.~\eqref{Eq:12b} follows that $\Gamma$ has a quadratic dependence on $\lambda$ with roots $\{0,\,-\gamma\}$ and a maximum $\Gamma_{\footnotesize\hbox{max}}=\gamma^2/4$ at $\lambda_{\footnotesize\hbox{max}}=-\gamma/2$. The system changes its stability properties when $\lambda$ crosses the vertical axis. Near the bifurcation point ($\lambda=0$), the stability of the mode is determined by the sign of $\Gamma$. Thus, the $n$-th mode is stable if
\begin{equation}
\label{Eq:17}
    \Gamma_n>0.
\end{equation}

From the previous results it is possible to predict the behaviour of the soliton bubble for different combinations of parameters. Interestingly, the condition of Eq.~\eqref{Eq:16} suggest not only that the first internal mode of the $\phi^4$ bubble can disappear for some combinations of parameters, but also that more than one internal mode can appear. Such new internal modes are also able to store energy from an external perturbation, such as the heterogeneity of Eq.~\eqref{Eq:06}, and can move with the bubble wall because they are localised around the soliton bubble. Thus, the soliton bubble can be regarded as a quasi-particle with intrinsic excitation modes. Moreover, Eq.~\eqref{Eq:17} suggest that such modes can also become unstable and lead to remarkable and complex phenomena, such as those detailed in the following sections.

\section{Bubble oscillations: over-damped and under-damped modes \label{Sec:OscillatingStates}}

\begin{figure}
    \centering
    \scalebox{0.8}{\includegraphics{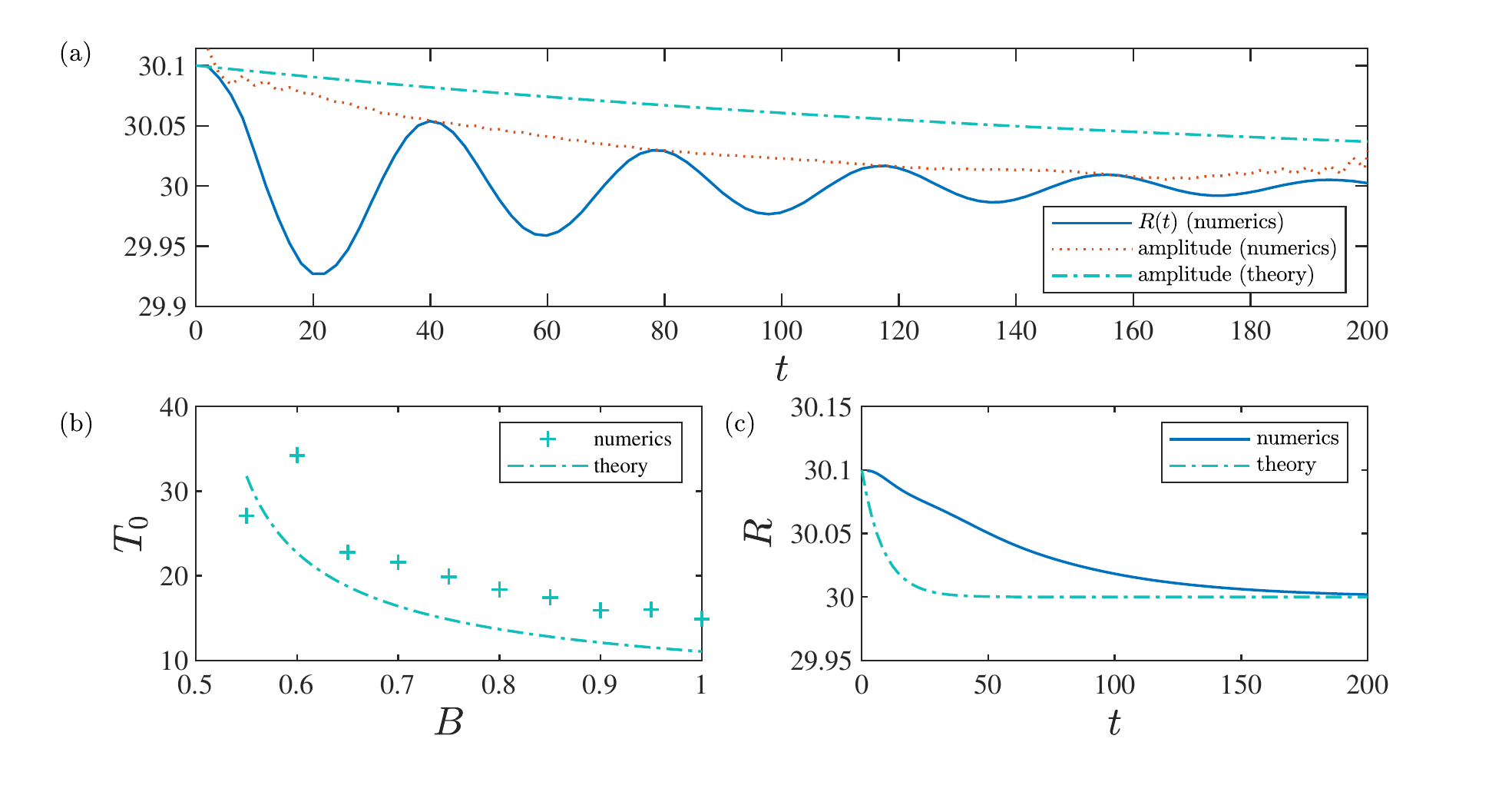}}
    \caption{Damped oscillations of the bubble-wall with initial radius $30.1$, for $r_o=30.0$, $\mu=1/2$, and $\eta=1$. \textbf{(a)} Typical outcome of under-damped oscillations in the wall position (blue solid line) for $B=0.55$ and $\gamma=0.01$. The amplitude of the oscillations computed through the Hilbert transform (red dashed line) is in good agreement with an exponential decay $R(t)=r_o\exp(\sigma t)$ with $\sigma=-\gamma/2$ and no fitted parameters (green dashed-dotted line). \textbf{(b)} Period of the under-damped oscillations as a function of parameter $B$ for $r_o=30.1$ and $\gamma=0.01$. Periods from numerical simulations (+) are in good agreement with the theoretical value of the period of the traslational mode (dashed-dotted line). The theoretical curve was obtained from Eqs.~\eqref{Eq:15a}, \eqref{Eq:15b} and \eqref{Eq:18a} for $n=1$ with no fitted parameters. \textbf{(c)} Position of the bubble-wall in the over-damped regime for $\gamma=0.6$ and $B=0.6$, using the theoretical model (green dashed-dotted line) and numerical simulations (blue solid line). }
    \label{fig:04}
\end{figure}

The first general consequence of the linear stability analysis is that stable modes can exhibit either under-damped oscillations or over-damped decay, whereas unstable modes can be only over-damped. Indeed, from Eq.~(\ref{Eq:12b}) follows that $\lambda=(-\gamma\pm\sqrt{\gamma^2-4\Gamma})/2$, which can be complex if $\gamma$ is small enough. This leads to a transient oscillatory behaviour around the bubble solution according to Eqs.~\eqref{Eq:10}.

Suppose that the $n$-th mode exists and is stable, i.e. $\Gamma_n>0$. If $\gamma^2<4\Gamma$, then $\lambda\in\mathbb{C}$ and the mode will perform \emph{under-damped} oscillations near the bifurcation point. The frequency and decay rate of such oscillations are given by
\begin{subequations}
\begin{equation}
\label{Eq:18a}
 \omega_n=\frac{1}{2}\sqrt{4\Gamma_n-\gamma^2},\end{equation}
 \begin{equation}
 \label{Eq:18b}
 \sigma=-\frac{\gamma}{2},
 \end{equation}
\end{subequations}
respectively. Several numerical simulations in the range $1/2<B<1$ using bubbles of initial radius slightly larger than $r_o=30.0$ have shown under-damped oscillations, as summarised in Fig.~\ref{fig:04}. Figure~\ref{fig:04}(a) shows a typical under-damped oscillatory behaviour in time of the bubble-wall position $R(t)$ when initially placed a small distance away from its equilibrium radius. The amplitude of the oscillations is computed by means of the Hilbert transform of $R(t)-r_o$ and is also shown in Fig.~\ref{fig:04}(a). These results are in agreement with an exponential decay $\exp(\sigma t)$ with $\sigma$ given by Eq.~\eqref{Eq:18b}. The frequency of oscillations is also computed by means of the fast Fourier transform of $R(t)$. Figure~\ref{fig:04}(b) shows the numerically calculated periods of oscillations for different values of $B$, which are in good agreement with the theoretical value $T_0=2\pi/\omega_0$. Notice that $T_o$ decays as $B$ increases, and increases indefinitely as $B\to1/2$. 

If $\gamma^2>4\Gamma_n$, then $\lambda\in\mathbb{R}$ and the mode decays with no oscillations near the bifurcation point. This corresponds to an \emph{over-damped} regime, where $\omega_n=0$ and the decay rate is
\begin{equation}
\label{Eq:19}
 \sigma=\frac{1}{2}\left(\sqrt{\gamma^2-4\Gamma_n}-\gamma\right).
 \end{equation}
In this latter regime, if the bubble wall is initially placed away from the stable equilibrium point, it will approach its equilibrium position asymptotically without oscillations, as depicted in Fig.~\ref{fig:04}(c). Finally, if the $n$-th mode is unstable, i.e. $\Gamma_n<0$, then $\lambda\in\mathbb{R}\forall\gamma\in\mathbb{R}$. Thus, the mode grows with no oscillations ($\omega_n=0$) near the bifurcation point with rate
\begin{equation}
\label{Eq:20}
\sigma=\frac{1}{2}\left(\sqrt{4\vert\Gamma_n\vert-\gamma^2}-\gamma\right).
\end{equation}
In this latter case, unstable modes do not exhibit oscillations.

\section{Vacuum decay in an effective heterogeneous potential \label{Sec:HeterogeneousVacDec}}

Section \ref{Sec:Bubbles} demonstrated that a localised heterogeneity can sustain a bubble. The heterogeneity deforms the effective energy landscape and stable phases become non-degenerated, with a true- and a false-vacuum state. This section shows that such heterogeneities may become a source of nonlinear instabilities that enhances the growth of bubbles of true vacuum. Eventually, these instabilities are strong enough to overcome the surface tension and the bubble expands indefinitely, thus triggering phase transitions and vacuum decay.

The fundamental mode ($n=0$) of the spectral problem \eqref{Eq:13}-\eqref{Eq:14} is
\begin{equation}
\label{Eq:21}
 \psi_0(r)=\mbox{sech}^{\Lambda}\left[B(r-r_o)\right],   
\end{equation}
where $\Lambda$ is given by Eq.~\eqref{Eq:15b} \cite{Flugge2012, Holyst1991, Gonzlez2008}. This mode exists $\forall B<+\infty$, following Eq.~\eqref{Eq:17}. However, the stability of the fundamental mode depends on parameter $B$. The critical condition $\Gamma_{0}=\Gamma_{0,c}:=0$ gives the threshold of instability of the fundamental mode, which is
\begin{equation}
    \label{Eq:22}
    B_c^{(0)}=\frac{1}{2}.
\end{equation}

\begin{figure}
    \centering
    \scalebox{0.411}{\includegraphics{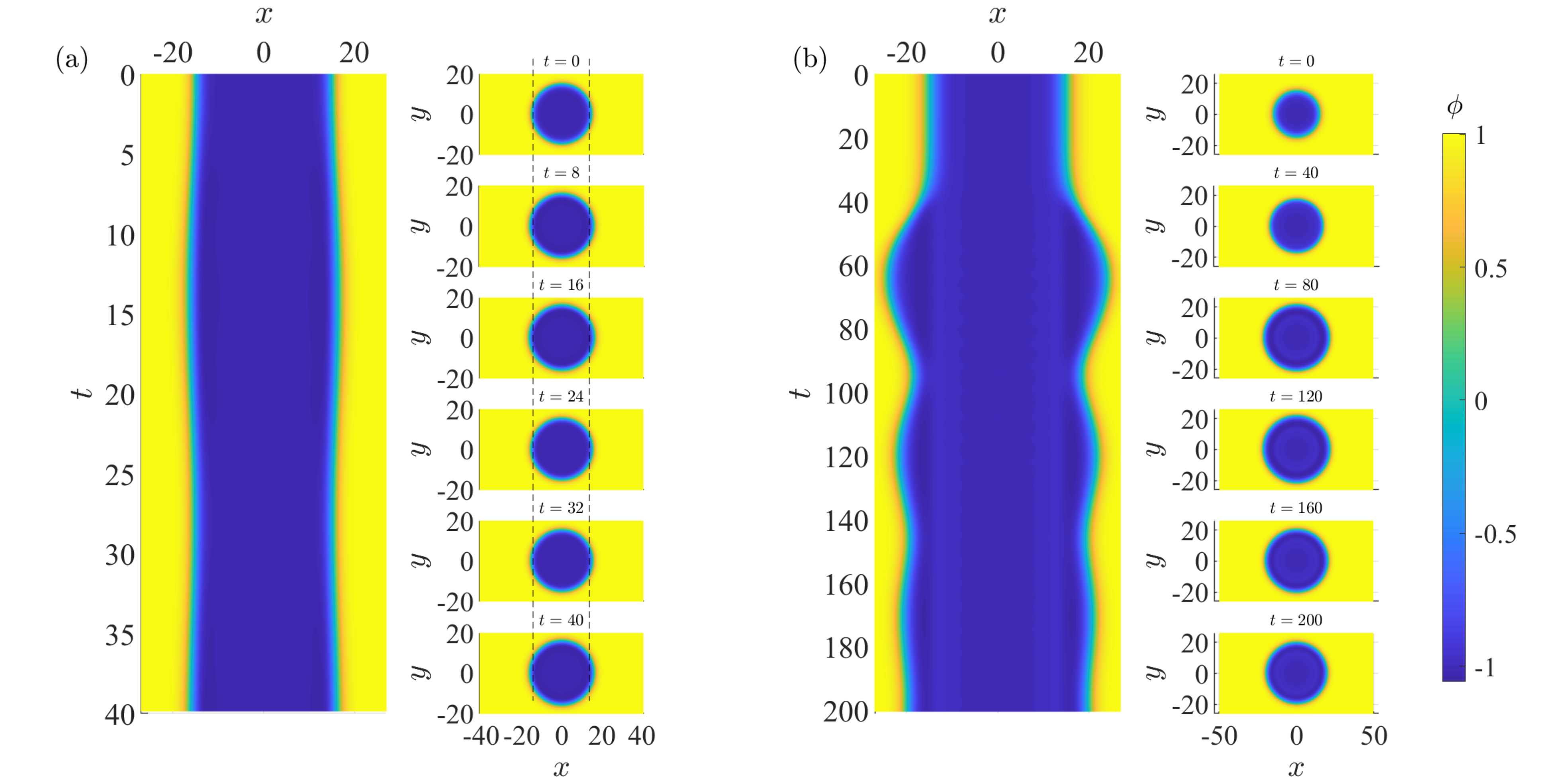}}
    \caption{\textbf{(a)} Stable bubble for $B=0.6$ and $r_o=15.0$. The system has two stable bound states. The insets show the snapshots of the bubble for the indicated values of the time. Vertical dashed lines indicate the initial position of the bubble wall for comparison. \textbf{(b)} Bubble oscillations when the fundamental mode is unstable for $B=0.4$. For both cases $\mu=1/2$, $\eta=1$ and $\gamma=0.01$.}
    \label{fig:05}
\end{figure}

The fundamental mode is stable ($\Gamma_0>0$) if $B>B_c^{(0)}$, and turns unstable if $B<B_c^{(0)}$. Notice that $\Lambda=2$ at the threshold of instability of the fundamental mode, which is explicitly given by
\begin{equation}
\label{Eq:23}
\psi_{0}(r)=\mbox{sech}^2\left[\frac{1}{2}(r-r_o)\right].
\end{equation}
 
For $r$ sufficiently large, the system possesses translational invariance and Eq.~\eqref{Eq:23} is the Goldstone mode. Such mode arises directly from the differentiation of bubble solution \eqref{Eq:05}. Given that the derivative operator is the generator of the translation group, this fundamental mode is a \emph{translational mode}. In the case of the driven bubbles shown in Figs.~\ref{fig:02}(c) and \ref{fig:03}, there is only one stable bound state for the given combination of parameters. Thus, the bubble is stable. Below the threshold of instability, i.e. for $B<B_c^{(0)}$, the bubble is expected to expand due to the instability of the translational mode.

If $B<B_*^{(1)}:=\sqrt{3}/2$, the first excited mode ($n=1$) exists and is given by \cite{Flugge2012, Holyst1991, Gonzlez2008}
\begin{equation}
\label{Eq:24}
 \psi_1(r)=\sinh[B(r-r_o)]\mbox{sech}^{\Lambda}[B(r-r_o)].
\end{equation}
The threshold of instability $B_c^{(1)}$ of this excited mode is obtained from Eq.~\eqref{Eq:17} with $n=1$, and gives
\begin{equation}
\label{Eq:25}
 B_c^{(1)}:=\left(\frac{11-\sqrt{117}}{8}\right)^{1/2}.
\end{equation}

The excited mode \eqref{Eq:24} exists and is stable if $B_c^{(1)}<B<B_*^{(1)}$, and turns unstable if $B<B_c^{(1)}$. Notice that the stable excited mode can coexist with the stable fundamental mode, since $B_c^{(1)}<B_c^{(0)}<B_*^{(1)}$. This is the case of the numerical simulations shown in Fig.~\ref{fig:05}(a), where the bubble wall performs stable oscillations due to the stability of modes $n=0$ and $n=1$.

\begin{figure}[b]
    \centering
    \scalebox{0.47}{\includegraphics{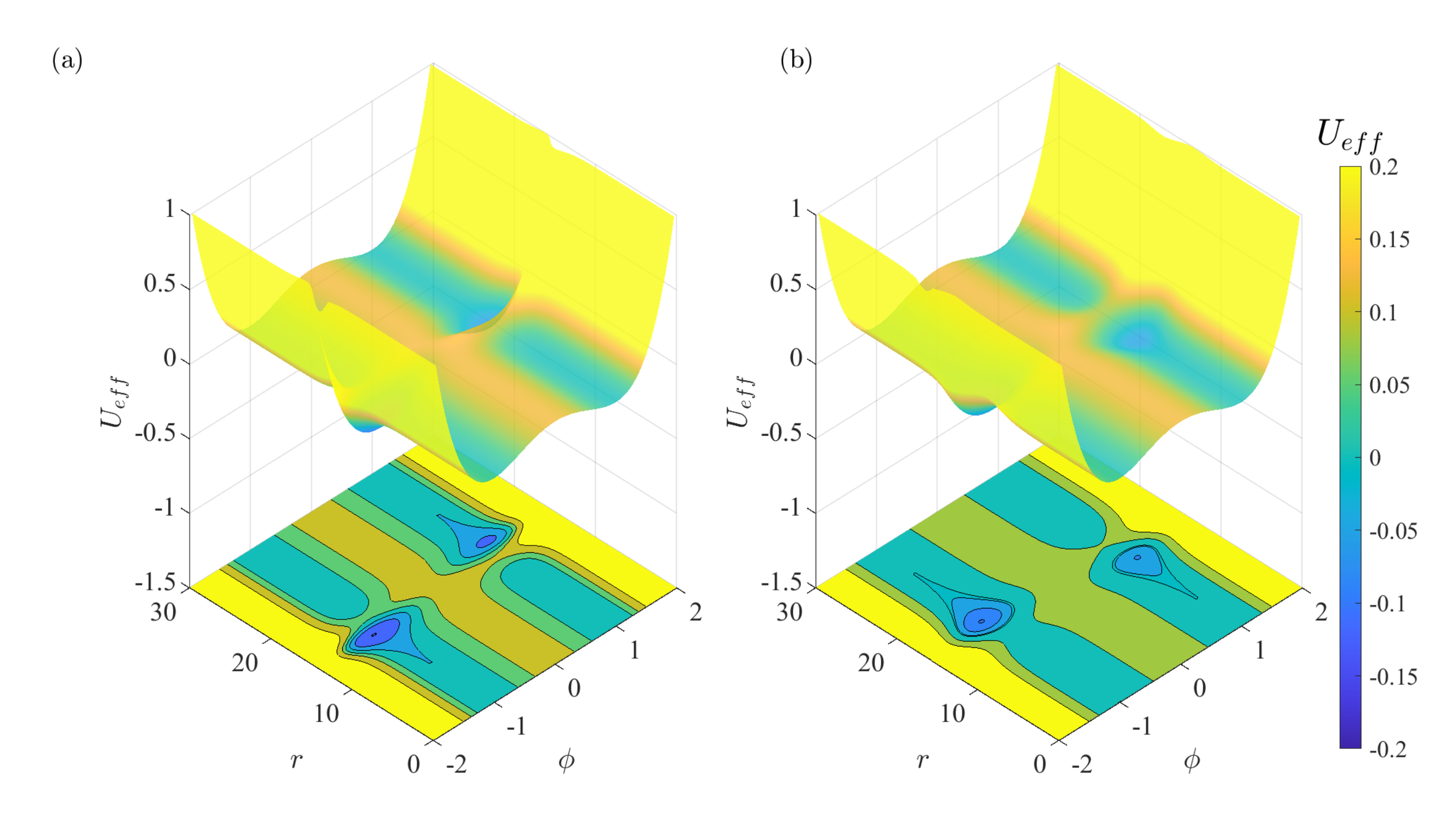}}
    \caption{Effective heterogeneous potential for \textbf{(a)} $B=0.6$ ($B>B_c^{(0)}$) and \textbf{(b)} $B=0.4$ ($B<B_c^{(0)}$). For both cases $r_o=15.0$, $\mu=1/2$ and $\eta=1$.}
    \label{fig:06}
\end{figure}

As $B$ approaches $B_c^{(0)}$, the period of oscillations of the fundamental mode increases towards infinity --see Fig.~\ref{fig:03}(b). Bellow the threshold of instability of the fundamental mode, the radius of the bubble will describe an exponential growth with the rate given by Eq.~\eqref{Eq:20} until nonlinearities become important and compensate the growth. This case is shown in the numerical simulations of Fig.~\ref{fig:05}(b), where there is an initial burst in the bubble area enhanced by the instability of the fundamental mode. This behaviour suggests the existence of a repulsive bubble-heterogeneity interaction. Indeed, the derivative of $f(r)$ at $r=r_*$ becomes negative for $B<B_c^{(0)}$. Thus, according to \eqref{Eq:07}, the point $r=r_*$ becomes unstable for the bubble wall. Eventually, the bubble area performs damped oscillations around a new equilibrium point, which is ruled by the compensation of the surface tension and the repulsion from the heterogeneity.

Figure \ref{fig:06} shows how the energy landscape is changed when crossing the threshold of instability of the fundamental mode. Compared to the case of Fig.~\ref{fig:03}(b), the potential wells are broader and are no longer symmetric. Now there is a true-vacuum state with $\phi=\phi_1$ and a false-vacuum state with $\phi=\phi_2$. The barrier that prevents false-vacuum to become true-vacuum is now significantly smaller than the barrier of Fig.~\ref{fig:03}(b). When crossing the threshold of instability, the false-vacuum potential well passes from the region $r\geq r_o$ [see Fig.~\ref{fig:06}(a)] to the region $r\leq r_o$ [see Fig.~\ref{fig:06}(b)]. Conversely, the true-vacuum potential well passes from the region $r\leq r_o$ to the region $r\geq r_o$. Thus, bubbles of the true vacuum state become larger below the threshold.

\begin{figure}
    \centering
    \scalebox{0.7}{\includegraphics{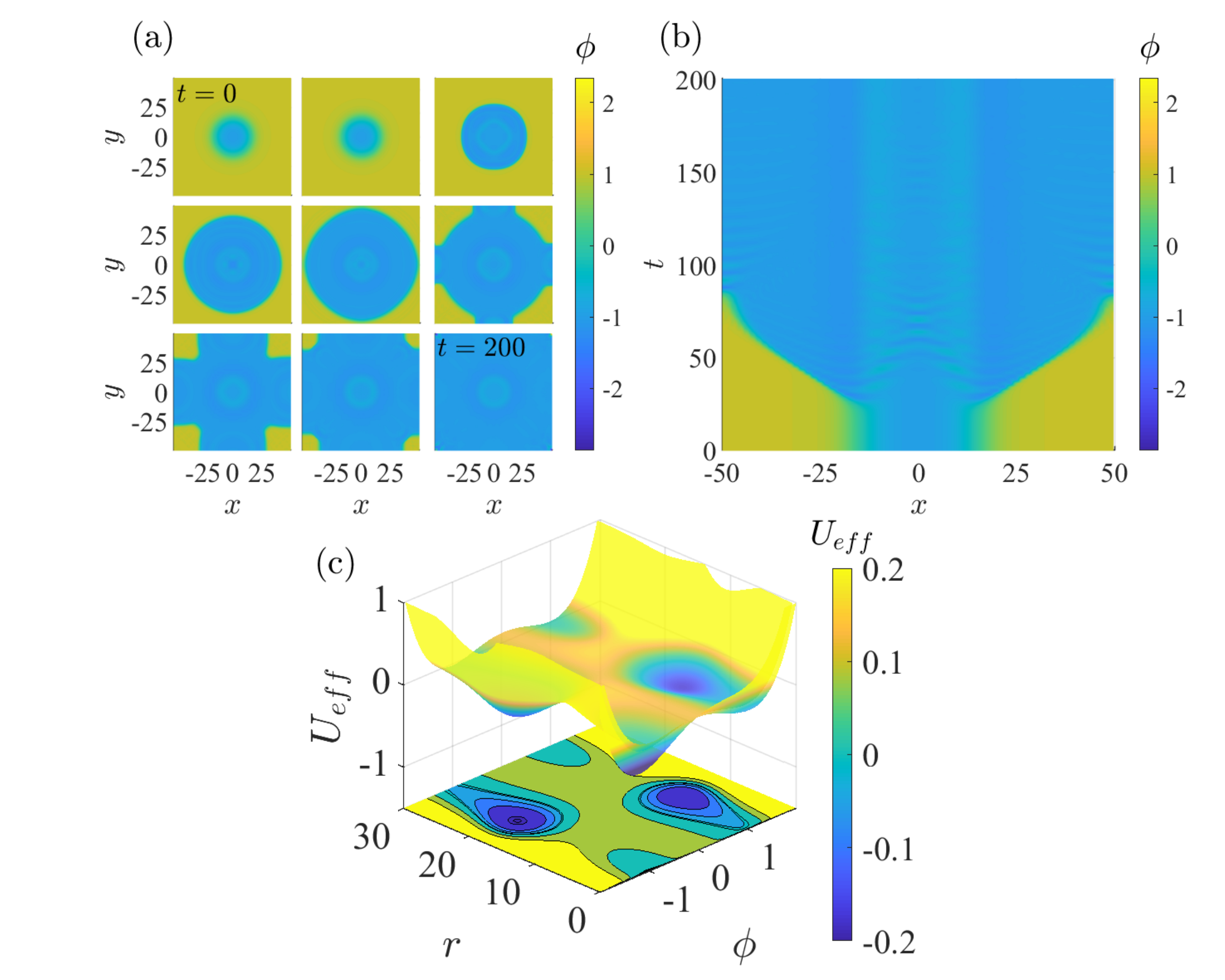}}
    \caption{Vacuum decay for $B=0.16$, $r_o=15.0$, $\mu=1/2$, $\eta=1$ and $\gamma=0.01$. \textbf{(a)} Snapshots from numerical simulations taken with steps  $\Delta t=25$. Time increases from left to right and from top to bottom. The initial and final values of time are indicated in the figure. \textbf{(b)} Time evolution of the $x$ profile of the true-vacuum bubble for $y=0$. \textbf{(c)} Energy landscape with a true vacuum state inside a broad potential well with $r\geq r_o$.}
    \label{fig:07}
\end{figure}

Further decreasing the value of $B$ in the interval $(B_c^{(1)},\,B_c^{(0)})$, the true-vacuum potential well becomes broader, allowing the existence of bubbles of true-vacuum with increasing radius. Thus, a natural question is whether vacuum decay can be enhanced by the instability of the translational mode. Indeed, Fig.~\ref{fig:07} shows a true-vacuum bubble expanding throughout the entire simulation space for $B=0.16$, converting false-vacuum into true-vacuum. The curvature of the bubble decreases as the radius of the bubbles increases, and the dynamics of the bubble is eventually dominated by the instability of the translational mode. Due to the finite computational domain in the numerical simulations of Fig.~\ref{fig:07}(a) and Fig.~\ref{fig:07}(b), the bubble eventually collides with the boundaries and some small-amplitude waves reflects towards the origin. These waves eventually dissipate in the presence of damping, resulting in a flat configuration with the space filled with phase $\phi_1$.

It is important to remark that not all bubbles of phase $\phi_1$ grow for the combination of parameters shown in Fig.~\ref{fig:07}. Figure~\ref{fig:07}(c) also evidences the existence of another potential well for bubbles of phase $\phi_1$ around $r=0$. Thus, true-vacuum bubbles with a relatively small radius ($r_o<r_*$) fall into this potential well and thus collapses towards the origin. This observation was confirmed by numerical simulations. In such a case, since $r_o<r_*$, both the repulsive interaction with the heterogeneity and the surface tension enhances the collapse of the bubble.

\section{Bubble nucleation inside a precursor bubble \label{Sec:Precursor}}

The first excited mode turns unstable if $B<B_c^{(1)}$, where $B_c^{(1)}$ is given by Eq.~\eqref{Eq:25}. As previously discussed, these excited modes are associated with shape-modes, which are intrinsic excitation modes of the bubble. Previous sections demonstrated that the instability of the translational mode produces bubble expansion or bubble collapse. However, more complex phenomena can occur when energy is stored in the internal (shape) modes of the bubble. Shape modes are responsible for variations in the kink-like profile of the bubble wall and may affect the global dynamics of the bubble. A phenomenon termed as \emph{shape-mode instability} occurs when a shape mode becomes unstable \cite{Gonzalez2002, Gonzalez2003, GarciaNustes2017, Marin2018, Castro-Montes2020}. In this scenario, the shape of the wall is no longer stable and the bubble breaks up.

\begin{figure}
    \centering
    \scalebox{0.54}{\includegraphics{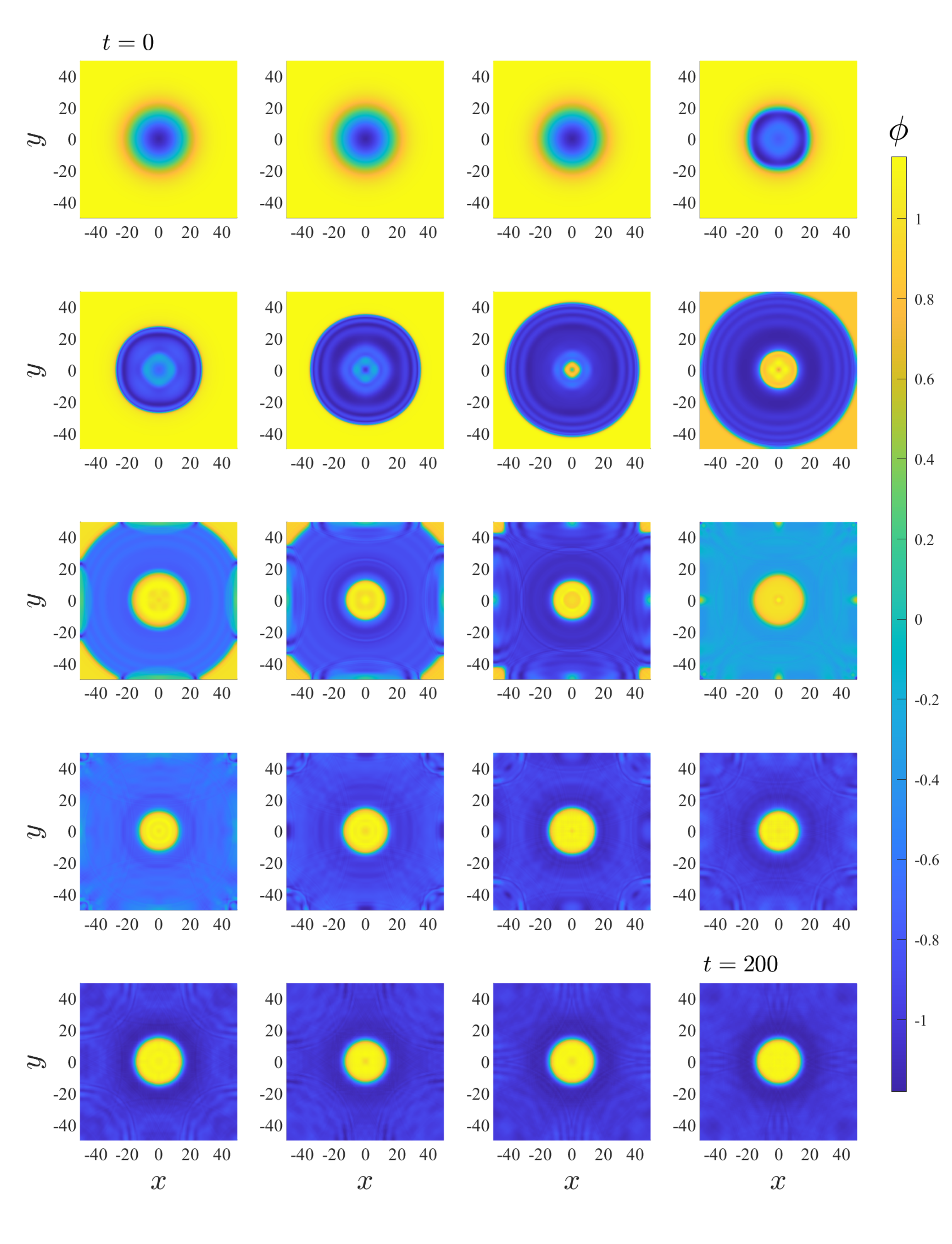}}
    \caption{Nucleation of a $\phi_2$-bubble inside an expanding $\phi_1$-bubble due to an internal-mode instability. Snapshots from numerical simulations taken with steps $\Delta t\simeq10.5$ for $B=0.1$, $\mu=1/2$, $\eta=1$, and $\gamma=0.01$. Time increases from left to right and from top to bottom.}
    \label{fig:08}
\end{figure}

Figure \ref{fig:08} shows the nucleation of a bubble of phase $\phi_2$ inside an expanding bubble of phase $\phi_1$ due to a shape-mode instability. Initially, the $\phi_1$-bubble expands due to the instability of the fundamental mode. However, the first internal mode of the bubble absorbs enough energy from the heterogeneity to become unstable and break the structure. Most of the absorbed energy is used for the nucleation of a new stable bubble of phase $\phi_2$ inside the precursor $\phi_1$-bubble. The remaining energy after the nucleation turns into small-amplitude radiative waves that dissipates in the presence of damping. The final configuration of the field is a bubble of phase $\phi_2$ sustained by the heterogeneity in a space filled with phase $\phi_1$.

\begin{figure}
    \centering
    \scalebox{0.5}{\includegraphics{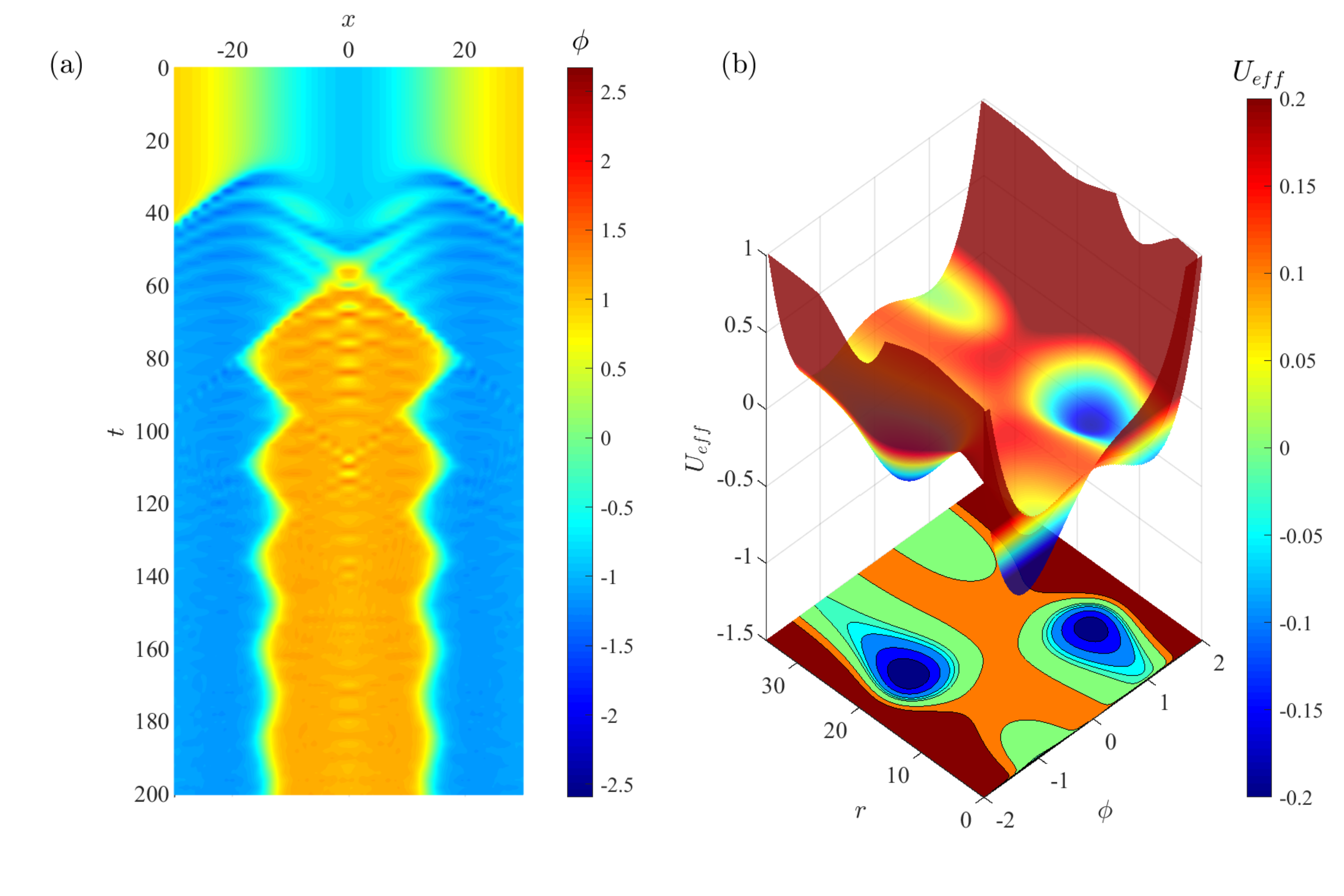}}
    \caption{Numerical simulations for $B=0.1$, $\mu=1/2$, $\eta=1$, and $\gamma=0.01$. \textbf{(a)} Spatiotemporal diagram at $y=0$ and \textbf{(b)} heterogeneous effective potential for the bubble breakup shown in Fig.~\ref{fig:08}.}
    \label{fig:09}
\end{figure}

Figure \ref{fig:09}(a) shows the evolution in time of the $x$-profile of the bubble breakup process. The oscillations of the internal modes are stronger when compared to previous cases, and a new $\phi_2$-bubble is generated from the origin. Following Eq.~\eqref{Eq:07}, the radius $r=r_*$ is a stable position for the wall of the newborn $\phi_2$-bubble, and an attractive interaction between its wall and the heterogeneity is on sight. Thus, the $\phi_2$-bubble expands and performs damped oscillations around the stable equilibrium radius $r=r_*$. At $t\to\infty$, the new bubble becomes stationary. The corresponding heterogeneous potential landscape is shown in Fig.~\ref{fig:09}(b). There is also a potential well for bubbles of phase $\phi_1$ around $r=0$, and now the stable phases $\phi_1$ and $\phi_2$ are slightly non-degenerated.

Further decreasing the value of $B$ below the threshold $B_c^{(1)}$, more excited modes can appear and store energy from the heterogeneity. Such modes can also become unstable for $B$ sufficiently small. The dynamics of the bubble become more complex in the presence of a mixture of stable and unstable internal modes.

\section{Bubble nucleation under thermal noise \label{Sec:Noise}}

Numerical simulations shown in Sections \ref{Sec:OscillatingStates}, \ref{Sec:HeterogeneousVacDec} and \ref{Sec:Precursor} demonstrated bubble dynamics and nucleation using the $\phi_1$-bubble of Eq.~\eqref{Eq:05} as the initial condition. The purpose of using such initial condition was to reproduce the conditions assumed in the linear stability analysis of Section \ref{Sec:LinearStability}, whose predictions have been confirmed by numerical results. Finally, this section shows that these results are robust to changes in the initial conditions, even in the presence of thermal fluctuations.

Consider the following driven-damped $\phi^4$ system
\begin{equation}
    \label{Eq:26}
 \partial_{tt}\phi-\nabla^2\phi+\gamma\partial_t\phi-\frac{1}{2}\left(1-\phi^2\right)\phi=f(r)+\tilde\xi(\mathbf{r},t),
\end{equation}
where $f(r)$ is given by Eq.~\eqref{Eq:06}, $\tilde\xi(\mathbf{r},t)$ is a space-time white noise source with $\langle\tilde\xi(\mathbf{r},t)\rangle=0$ and  $\langle\tilde\xi(\mathbf{r},t)\tilde\xi(\mathbf{r}',t')\rangle=2k_BT\delta(t-t')\delta(\mathbf{r}-\mathbf{r}')$. Here, $T$ has the interpretation of temperature and $k_B$ is the Boltzmann's constant \cite{Lythe2019}. The $\phi^4$ potential $U(\phi)$ has two degenerated phases $\phi_1=-1$ and $\phi_2=1$ separated by a barrier at $\phi_0=0$. Both the heterogeneity $f(r)$ and the thermal noise can deform such potential and break the degeneracy of the minima of $U(\phi)$. To check the robustness of the results of this article on initial conditions, consider the following initial configurations of the field $\phi$ given by a random distribution of values in space,
\begin{equation}
    \label{Eq:27}
    \phi(\mathbf{r},t=0)=\phi_i+\delta\tilde\epsilon(\mathbf{r}),
\end{equation}
where $i=0,1,2$ labels the stable phases of potential $U(\phi)$, $\tilde\epsilon$ is a delta-correlated space dependent random variable with $\langle\tilde\epsilon\rangle=0$, and $\delta=0.0005$ is the noise strength.

\begin{figure}
    \centering
    \scalebox{0.47}{\includegraphics{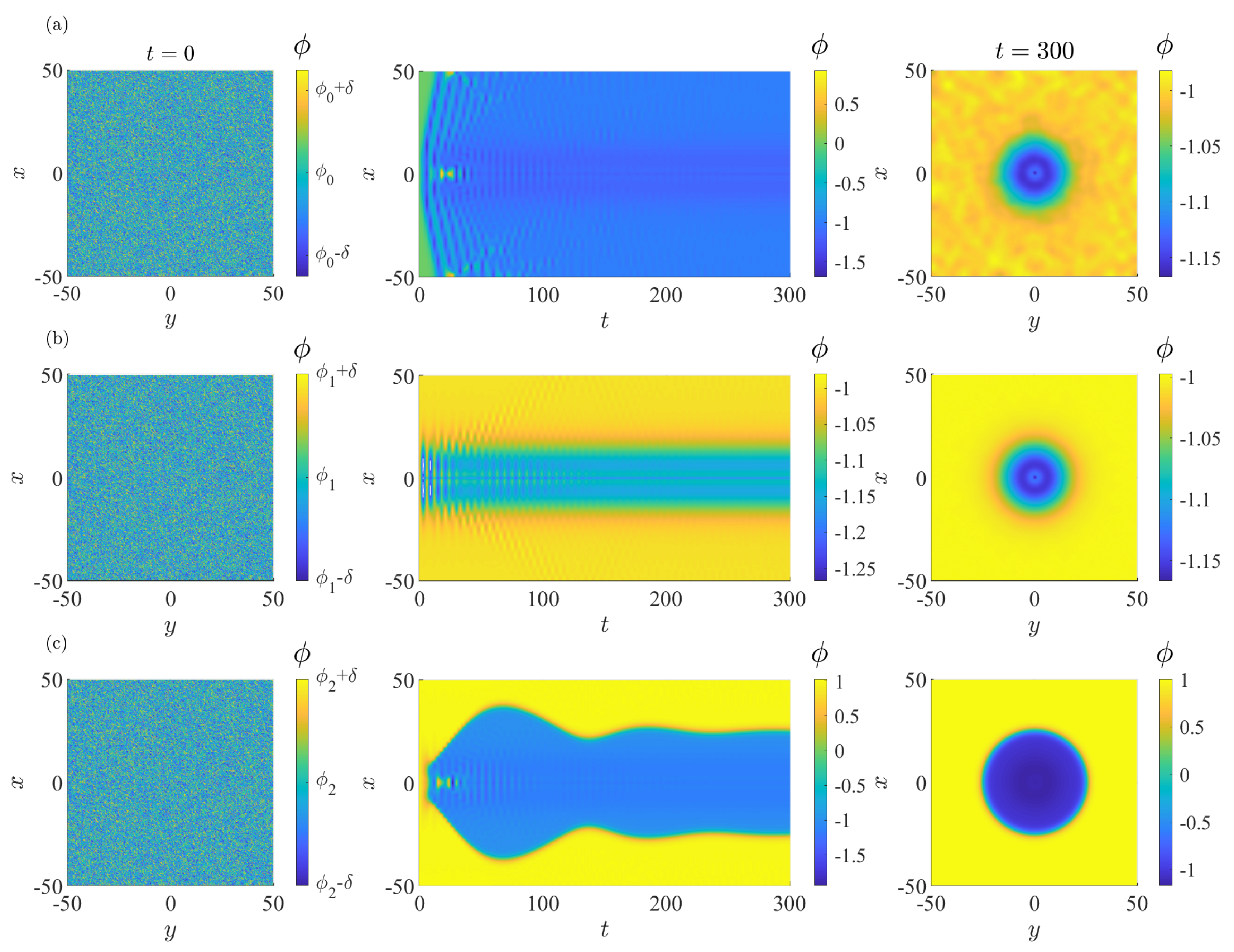}}
    \caption{Evolution of random initial conditions under thermal fluctuations for $\delta=0.0005$, $B=0.1$, $\mu=1/2$, $\eta=1$, and $\gamma=0.01$. Initial conditions for each case are shown in the left column. The central column shows the spatiotemporal evolution of the $x$-profile of $\phi$ at $y=0$. The right column shows the field configuration at the end of the simulation. \textbf{(a)} If the initial condition is a noisy distribution around the barrier of the $\phi^4$ potential, the system decays towards the phase $\phi_1$. \textbf{(b)} If the initial condition is noisy around the phase $\phi_1$, the system remains in the corresponding potential well. \textbf{(c)} If the initial condition is noisy around the phase $\phi_2$, a $\phi_1$-bubble is nucleated around the heterogeneity.}
    \label{fig:10}
\end{figure}

Figure~\ref{fig:10}(a) shows the results from numerical simulations for $i=0$ and the given values of parameters. In this case, the initial condition of Eq.~\eqref{Eq:27} is a small-amplitude random distribution of values around the unstable equilibrium state $\phi_0$. After a transient of waves emitted from the heterogeneity and reflected from the boundaries, the system finally decays to the potential well of phase $\phi_1$ throughout all space. This is expected given that $\phi_1$ is the true-vacuum state. At $t=300$ the field is in a noisy configuration inside the potential well corresponding to the phase $\phi_1$. Notice that the linear response of the field to the ring-shaped heterogeneity is appreciable at the end of the simulation.

For the case $i=1$, Eq.~\eqref{Eq:27} is a small-amplitude random distribution of values around the stable equilibrium phase $\phi_1$. Given that the noise strength is small and unable to induce a phase transition by itself, the field $\phi$ remains in such potential well, as demonstrated in Fig.~\ref{fig:10}(b). Noise dissipates fast in this case, and the final configuration is the linear response of the field to the heterogeneity, almost free of noise.

For the case $i=2$, the initial condition is a random distribution around the meta-stable phase $\phi_2$, and the outcome is different. The heterogeneity seeds a phase transition, and a bubble of phase $\phi_1$ is nucleated around the heterogeneity. Figure~\ref{fig:10}(c) shows how the area of the $\phi_1$-bubble initially grows and performs damped oscillations around the equilibrium radius. In conclusion, a heterogeneity under the effects of thermal noise can nucleate bubbles in this two-dimensional $\phi^4$ system.

\section{Conclusions \label{Sec:Conclusions}}

In summary, this article investigated a two-dimensional $\phi^4$ model exhibiting bubble nucleation around heterogeneities. Bubbles are studied as radially symmetric heteroclinic solutions interpolating two stable phases of the underlying potential. Through the solution of an inverse problem, a nonlinear field theory with exact solutions is proposed. The dynamics and stability of soliton-bubbles in such a theory is investigated under the influence of the heterogeneity. Numerical and analytical results have shown that bubbles can be nucleated and sustained by the heterogeneity, even in the presence of thermal noise. The model predicts the formation of oscillating bubbles, vacuum decay, and the nucleation of bubbles inside a precursor expanding-bubble, depending on the combination of parameters associated with topological properties of the heterogeneity. Numerical simulations have shown great agreement with analytical results. These results may be useful in the study of simple models of vacuum decay seeded by heterogeneities, as well as the nucleation and stability of localised structures in the Universe with a long life-time containing new physics.

\begin{acknowledgments}
The author thanks M. Ahumada for her advice on the article. This work was partially supported by Universidad de Santiago de Chile through the POSTDOC\_DICYT project number 042031GZ\_POSTDOC and by ANID FONDECYT/POSTDOCTORADO/3200499.
\end{acknowledgments}



\providecommand{\noopsort}[1]{}\providecommand{\singleletter}[1]{#1}%

\end{document}